\documentstyle[aps,12pt,psfig]{revtex}
\begin{document}
\tightenlines
\title{\Large Factorization approach for barotropic FRW model with a 
cosmological constant}
\author{J. Socorro, M.A. Reyes and F.A. Gelbert} 
\address{Instituto de F\'{\i}sica de la Universidad de Guanajuato,\\
Apartado Postal E-143, C.P. 37150, Le\'on, Guanajuato, M\'exico}

\maketitle
\widetext

\begin{abstract}
We apply the technique of standard supersymmetric factorization for any
q factor ordering in the Wheeler-DeWitt (WDW) equation for barotropic 
Friedmann-Robertson-Walker (FRW) minisuperspace model, including the 
cosmological term. The resulting wave functions of the universe are
exhibited as one-parameter families, which for particular values of
the $\gamma$ parameter of the barotropic model are exactly found. 
\end{abstract}

\noindent {\rm PACS numbers: 02.30.Hq, 04.20.Jb, 04.40.Nr, 98.80.Hw.}

\section{Introduction}
The study of eigenvalue problems associated with second-order differential
operators found a renewed impulse with the application of the factorization
technique and its generalizations \cite{mielnik,nieto,ferna,cooper},
which was fundamented on the inverse scattering method of Gelfand
and Levitan \cite{gel}.

The use of the strictly isospectral scheme based on the general
Riccati solution\cite{mielnik,nieto,ferna,cooper}, has been applied
from classical and quantum physics\cite{mielnik}  to relativistic models
\cite{lanl}. This  technique has been known since about a decade in 
one-dimensional supersymmetric quantum mechanics (SUSY-QM) and usually 
requires 
nodeless, normalizable states of
Schr\"odinger equation. However, Pappademos, Sukhatme, and Pagnamenta
\cite{sukatme1} showed thah the strictly isospectral construction
can also be performed on non-normalizable states.

SUSY-QM may be considered an equivalent formulation of the Darboux 
transformation method, which is  well-known in mathematics from the original 
paper of Darboux \cite{darboux}, book by Ince \cite{ince}, and book by 
Matveev and Salle \cite{matveev}, where the method is widely used in the 
context
of the soliton theory. An essential ingredient of the method is a particular
choice of a transformation operator \cite{gel} in the form of a differential
operator \cite{sanson} which intertwines two hamiltonians and relates their
eigenfunctions. When this approach is applied in quantum theory it allows 
one to generate a huge family of exactly solvable local potential starting
with a given exactly solvable local potential \cite{cooper}.

In nonrelativistic one-dimensional supersymmetric quantum mechanics, the
factorization technique was applied to the $\rm q=0$ factor ordered WDW 
equation corresponding to the FRW cosmological models without matter field 
\cite{RoSo}, where a one-parameter class of strictly isospectral cosmological
FRW solutions was exactly found, representing the wave functions of the 
universe for that case. Also, in Ref. \cite{Roso1} a  one-parameter family of 
closed, radiation-filled FRW quantum universe for any $\rm q$ factor order 
was found. 

In this work, we extend the application of the factorization technique to
the case $\rm q \ne 0$ factor ordering in the WDW equation for the
barotropic FRW minisuperspace model, including a cosmological term.

The work is organized as follows. In section II we describe the 
supersymmetric factorization using the hamiltonian of the barotropic FRW 
cosmological model, including any factor ordering in the 
hamiltonian operator.
 In section III, we present the general solution to the Riccati equation
obtained when supersymmeterizing this hamiltonian,  as a strictly isospectral 
one-parameter family of 
both FRW cosmological potential and wave function of the model. Section IV is 
devoted to explicitely obtain the exact  cosmological solutions of the WDW 
equation for some typical values  of the $\gamma$ and $\kappa$ parameters
in the literature. We give final remarks in  section V.

\section{Supersymmetric factorization scheme}
We start with the hamiltonian  that appears in the study of the barotropic
FRW cosmological model, with cosmological constant (Wheeler-DeWitt equation), 
\begin{equation}
\rm \hat {\cal H}  |\Psi \rangle = \frac{1}{24A}\left[-\frac{d^2}{dA^2} 
+144\kappa A^2 
+48 \Lambda A^4 -384\pi G  M_\gamma A^{-3\gamma +1}\right] |\Psi
\rangle=0,
\label{hamiltonian1}
\end{equation}
where $\rm A$ is the scale factor, $\kappa$ is the curvature index of the 
universe ( $\rm \kappa=0,+1,-1$ plane, close and open, respectively),
$\Lambda$ the cosmological term and $\gamma$ the parameter that describes 
the state equation of the fluid.

In principle, the order ambiguity in equation 
(\ref{hamiltonian1}) should be taken into account. This is quite a 
difficult problem to be treated in all its generality, since the 
hamiltonian operator in (\ref{hamiltonian1}) must be written in 
a very general form in order to take into account all possible orders.
For  the semi-general Hartle-Hawking factor ordering\cite{HH}, we have 
\footnote{There are other possibilities depending on 
different considerations on the operators, see refs. \cite{za1,li1}}

\begin{equation}
\rm A^{-1} \frac{d^2 \Psi }{d A^2} \rightarrow A^{-1+q} \frac{d }{d A} A^{-q} 
\frac{d \Psi}{d A}= 
A^{-1}\left( \frac{d^2 \Psi}{d A^2} - q A^{-1} \frac{d \Psi}{dA} \right ),
\label{modi}
\end{equation}
where the real parameter $\rm q$ measures the ambiguity in the factor ordering.
In this approach, the Wheeler-DeWitt equation can be written as follows
\begin{equation}
{\cal H}_0 \Psi= \rm 
 - A\frac{d^2\Psi}{dA^2} + q \frac{d \Psi}{dA} + V(A)\Psi=0,
\label{WDW}
\end{equation}
where 
\begin{equation}
\rm V(A)= 144 \kappa A^3+48 \Lambda A^5 - 384 \pi G M_\gamma A^{-3\gamma + 2} .
 \label{pote}
\end{equation}

Our interest here is to study  the quantum solutions of equation 
(\ref{WDW}) when using the supersymmetric factorization scheme. As we shall
show, the solutions can be exactly found for particular values of the 
$\gamma$ parameter and the factor ordering parameter q. 

Consider the equation (modified WDW)
\begin{equation}
\rm H^+ \Psi= - A^{-2q}\frac{d^2\Psi}{dA^2} +q A^{-1-2q} \frac{d \Psi}{dA} + 
A^{-1-2q}V(A)\Psi=0.
\label{WDW-modi}
\end{equation}
It is easy to show that the first  order  differential operators
\begin{eqnarray}
{\cal A}^+ &=& {\rm  - A^{-q} \frac{d}{dA} + W(A)}, \label{a+}\\
{\cal A}^- &=& {\rm   A^{-q} \frac{d}{dA} + W(A)}, \label{a-}
\end{eqnarray}
where $\rm W$ plays the  role of a superpotencial function, factorize
the hamiltonian (\ref{WDW-modi}) as 
\begin{equation}
\rm H^+= {\cal A}^+ {\cal A}^- \, .
\label {h+}
\end{equation}
The potential term $\rm V(A)$ is related to the superpotential function
$\rm W(A)$ via the Ricatti equation\footnote{Hereafter we shall use 
$\rm W_\gamma$ to explicitely denote the superpotential functions depend
on the actual choice of the $\gamma$ parameter, and we shall call 
$\rm V_+$ to $\rm V$ in (\ref{WDW-modi}).}

\begin{equation}
\rm V_+(A,\gamma)= A^{1+2q} W^2_\gamma - A^{1+q} \frac{dW_\gamma}{dA}.
\label{ricatti}
\end{equation}

Making the transformation 
\begin{equation}
\rm W_\gamma= -A^{-q} \frac{u^{\prime}_{\gamma}}{u_\gamma},
\label{superpotential}
\end{equation}
where the $\prime$ means $\rm \frac{d}{dA}$,  
(\ref{ricatti}) is transformed into the original hamiltonian applied to the
function $\rm u_\gamma$, which means
that the superpotential function is know once we have a solution to the
original WDW equation. If we choose the ordinary factor ordering $\rm q=0$, we 
recover the relation found in \cite{RoSo}.

In the supersymmetric factorization scheme, $\rm V_-$, 
the partner superpotential of  $\rm V_+$, is obtained by performing
the product

\begin{equation}
\rm H^-  = {\cal A}^- {\cal A}^+ , \qquad H^-f(A)=0,
\label{h-}
\end{equation}
where $\rm f$ is the wave function related at the hamiltonian $\rm H^-$.
Then, the isospectral potential to $\rm V_+(A,\gamma)$ is
\begin{equation}
\rm V_-(A,\gamma)=A^{1+2q} W^2 + A^{1+q} W^{\prime}_\gamma=
V_+(A,\gamma)+ 2 A^{1+q} W^{\prime}_\gamma .
\label{iso} 
\end{equation}

In addition, the solution $\rm f(A)$ is found using the fact that 
${\cal A}^+ {\cal A}^- u_\gamma=0$. Multipliying both sides of Eq.
(\ref{iso}) by ${\cal A}^-$, 
we have that $\rm {\cal A}^- {\cal A}^+ \left( {\cal A}^- u \right)=
 H^- \left( {\cal A}^- u \right)=0$;
then $\rm f(A)= {\cal A}^-\, u$. In this way, the structure for $\rm f(A)$ 
becomes
\begin{equation}
\rm f(A)= W_\gamma u_\gamma + A^{-q} u^{\prime}_\gamma.
\end{equation}

Hence, knowing the superpotential function we can find the wave functions of
the partner hamiltonian $\rm H^-$. However, this is not the most general
solution, as  will be shown in the following section.

\section{general solution}
The general solution to the Ricatti equation (\ref{iso}) is found when we 
propose the
following scheme \cite{mielnik,cooper}
\begin{equation}
\rm V_-(A,\gamma)= A^{1+2q} \hat W^2 +A^{1+q} \hat W^\prime \equiv
 A^{1+2q} W^2_\gamma +A^{1+q}  W^{\prime}_\gamma,
\label{general-potential}
\end{equation}
which by choosing 
\begin{equation}
\rm \hat W \equiv W_\gamma + \frac{1}{y_\gamma},
\label{general}
\end{equation}
leads to a Bernoulli equation for $\rm y_\gamma$,
\begin{equation}
\rm y^{\prime}_\gamma - 2 W_\gamma A^q \,y_\gamma= A^q, 
\end{equation}
whose solution is
\begin{equation}
\rm y_\gamma (A)=u_\gamma^{-2} \left[ I_\gamma+ \lambda \right],
\end{equation}
where $\rm I_\gamma(A)= \int_0^A x^q u_\gamma^2 dx$.

In this way, (\ref{general}) can be written as
\begin{equation}
\rm \hat W(A)=W_\gamma + \frac{u_\gamma^2}{I_\gamma + \lambda},
\end{equation}
and the entire family of bosonic potentials can be built as
\begin{equation}
\rm \hat V_+(A,\gamma, \lambda)=A^{1+2q} \hat W^2(A,\gamma,\lambda) 
- A^{1+q} \hat W^\prime(A,\gamma,\lambda),
\end{equation}

\begin{eqnarray}
{\rm \hat V_+(A,\gamma,\lambda)}&=&{\rm  V_- - 2 A^{-q} \hat W^\prime }\\
&=& {\rm V_+(A,\gamma) - 4\frac{A^{1+q} u_\gamma u_{\gamma}^\prime}{I_\gamma 
+\lambda} + 2\frac{ A^{1+2q} u_\gamma^4}{(I_\gamma + \lambda)^2} }.
\label{iso-pote}
\end{eqnarray}

Finally, 
\begin{equation}
\rm \hat u_\gamma \equiv  g(\lambda) \frac{ u_\gamma}{ I_\gamma + \lambda},
\label{solu}
\end{equation}
is the isospectral solution of the Schr\"odinger equation (\ref{WDW})
for the new 
family potential (\ref{iso-pote}), with the condition on the function 
$\rm g(\lambda)= \sqrt{\lambda(\lambda+1)}$, though in the limit
\begin{equation}
\rm \lambda \to \pm \infty \qquad g(\lambda) \rightarrow \lambda \qquad and 
\quad \hat u_\gamma \rightarrow u_\gamma.
\end{equation}

This $\rm \lambda$ parameter is included not for factorization reasons, because
the wave functions in quantum cosmology are still nonnormalizable, but as
decoherence parameter embodying a sort of quantum cosmological dissipation
(or damping) distance.

 The WDW (\ref{WDW}) equation has 
particular solutions for the $\gamma$ parameter and different universes,
 which we shall explore in the following section.
\section{Solution of the  WDW equation for particular $\gamma$}

Quantum solutions can be readily found for particular choices of the
$\gamma$ parameter. Here, we list some of them.

\begin{enumerate}
\item{} $\gamma=-1$ corresponds to inflationary scenary.
For this stadium (\ref{WDW}) can be written

\begin{equation}
\rm A u^{\prime \prime}_{-1} -q u^{\prime}_{-1} + 
144 A^3 \left( m^2 A^2 -\kappa)   \right) u_{-1}=0,
\end{equation}
where $\rm m^2=- \frac{\Lambda}{3} + \frac{8}{3}\pi G M_{-1}$. 

Now, we have three possible cases in our analysis:

\noindent a)  $\rm m^2 > 0$, 

The differential equation for this subcase is

\begin{equation}
\rm A u^{\prime \prime}_{-1} -q u^{\prime}_{-1} + 144\kappa A^3
\left( m^2 A^2 - \kappa  \right)  u_{-1}=0,
\end{equation}
which, after the consecutive substitutions $\rm v=m^2 A^2 - \kappa$, 
$\rm u_{-1}=v^{1/2} y(v)$, and $\rm z=\frac{4}{m^2} v^{3/2}$, yields an
 ordinary Bessel equation for $\rm q=1$, with solution 
\begin{equation}
\rm u_{-1}=\left( m^2 A^2 - \kappa  \right)^{\frac{1}{2}}\left[a_0 
J_{\frac{1}{3}}(z) + b_0 J_{-\frac{1}{3}} (z)  \right] ,
\label{26}
\end{equation}
 where 
$ z=\frac{4}{m^2}\left[m^2 A^2 - \kappa\right]^{3/2}$,
and  $\rm a_0$ and $\rm b_0$ are superposition constants.

\begin{figure}[ht]
\centerline{\psfig{file=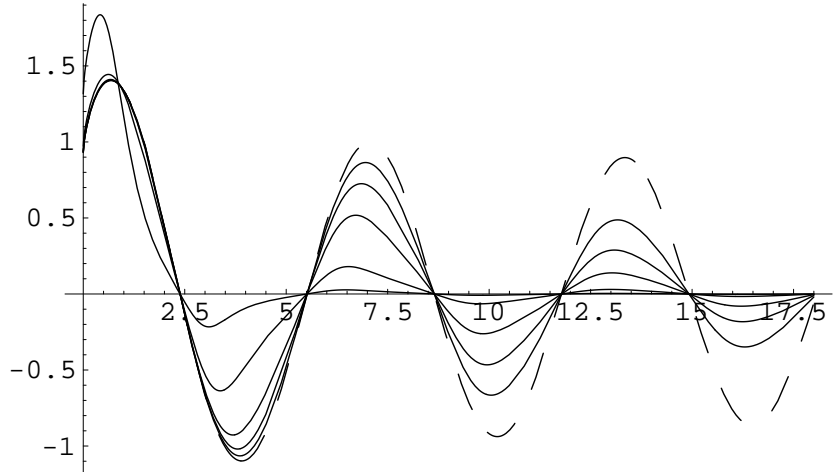,width=11.cm} }
\caption{The isospectral wave function  $\rm \hat u_{-1}$  (\ref{solu}) for 
$\rm m^2=4$, $\rm a_0= b_0=1$. Curves corresponding to increasing $\lambda$ 
(solid line, $\lambda=1,11,61,161,411$) path $\rm u_{-1}$ (dashed),  eq. 
(\ref{26}).}
\label{u-1}
\end{figure}

\noindent b) $\rm m^2 < 0$, 

The differential equation for this subcase is

\begin{equation}
\rm -A u^{\prime \prime}_{-1} +q u^{\prime}_{-1} + 144\kappa A^3
\left( |m^2| A^2 + \kappa  \right)  u_{-1}=0,
\end{equation}
being the solution the modified Bessel equation for $\rm q=1$ 
\begin{equation}
\rm u_{-1}=\left( |m^2| A^2 + \kappa  \right)^{\frac{1}{2}}
\left[a_1 I_{\frac{1}{3}}(z)
+ b_1 K_{\frac{1}{3}} (z)  \right] , 
\end{equation}
 where 
$z=\frac{4}{|m^2|}\left[|m^2| A^2 + \kappa\right]^{3/2}$,
and $\rm a_1$ and $\rm b_1$ are superposition constants.

\noindent c) For $\rm m^2=0$, 
and the differential equation for this situation is

\begin{equation}
\rm -A u^{\prime \prime}_{-1} +q u^{\prime}_{-1} + 144\kappa A^3  u_{-1}=0,
\end{equation}
which, for $\kappa=1$ has as solutions the modified Bessel functions
of order $\rm \nu=\frac{1+q}{4}$
\begin{equation}
\rm u_{-1}=A^{2\nu}\, \left[A_0 I_\nu(6A^2)
+ B_0 K_\nu (6A^2)  \right], 
\end{equation}
where $\rm A_0$ and $\rm B_0$ are superposition constants, whereas for
$\rm \kappa=-1$, the solutions become  the ordinary Bessel functions

\begin{equation}
\rm u_{-1}=A^{2\nu} \, \left[A_1 J_\nu(6A^2)
+ B_1 Y_{\nu} (6A^2)  \right], 
\end{equation}
where $\rm A_1$ and $\rm B_1$ are superposition constants.

\item{} Dust era, $\gamma=0$, and plane universe, $\kappa=0$:

By use of  the transformations $\rm z=\frac{8}{3}\sqrt{3\Lambda}\, A^3$ and
$\rm u_0=e^{-z/2} w(z)$, we find the hypergeometric differential equation
for $\rm w(z)$
\begin{equation}
\rm z \frac{d^2 w}{dz^2} + (\alpha - z) \frac{dw}{dz} -n w=0,
\end{equation}
where $\rm n=\frac{2-q}{6} - \frac{16\pi G M_0}{\sqrt{-3\Lambda}} $
and $\rm \alpha =\frac{2-q}{3}$.
Therefore, the independent solutions are \cite{gra}
\begin{eqnarray}
{\rm w_1}& =& {\rm _1F_1(n,\alpha;z)  }, \\ 
{\rm w_2}& =& {\rm z^{1-\alpha} {_1F_1}(n-\alpha +1,2-\alpha;z)  },
\label{ws} 
\end{eqnarray}
where $\rm _1F_1$ is the degenerate hypergeometric function. In this way, the
solution for this case become
\begin{equation}
\rm  u= e^{-z/2} \left[A_2 w_1(z) + B_2 w_2(z)\right],
\end{equation}
where $\rm A_2$ and $\rm B_2$ are superposition constants.

\item{} Stiff fluid, $\gamma=1$ and plane universe

The WDW equation for this stadium is written as
\begin{equation}\rm 
A u_1^{\prime\prime}- q u_1^\prime + 48\left(- \Lambda A^5 + 8\pi G
M_1 A^{-1} \right) u_1=0,
\end{equation}
whose solutions become
\begin{equation}
\rm  u=A^{\frac{1+q}{2}} Z_\mu \left(\frac{4\sqrt{-3\Lambda}}{3}A^3 \right) ;
\qquad
with \quad \mu=\frac{1}{3}\sqrt{(\frac{1+q}{2})^2-384 \pi G M_1},
\end{equation}
where $\rm Z_\mu$ is a generic Bessel function with  real or 
imaginary order  $\mu$\cite{imaginary}.

In this case, the exact expression of the solutions will depend on the signs 
of the cosmological constant and the parameter $\mu$.

\noindent i) $\mu$ real and $\Lambda >0$, the function $\rm Z_\mu$ become 
the modified Bessel functions,
either $\rm I_\mu$ or $\rm K_\mu$, depending on the boundary conditions.

\noindent ii) $\mu$ real and $\Lambda <0$, the functions $\rm Z_\mu$ turn 
into the ordinary Bessel function,
either $\rm J_\mu$ or $\rm Y_\mu$; depending on the boundary conditions.

\noindent iii) $\Lambda >0$ and $\mu$ pure imaginary, the functions 
$\rm Z_\mu$ become the modified Bessel functions  of pure imaginary order
\cite{imaginary}, either $\rm I_\mu$ or $\rm K_\mu$, depending on 
the boundary conditions.

\noindent iv) $\Lambda <0$ and $\mu$ pure imaginary, the functions 
$\rm Z_\mu$ turn into the ordinary Bessel functions of pure imaginary order 
\cite{imaginary}, either $\rm J_\mu$ or $\rm Y_\mu$, depending on 
the boundary conditions.
\end{enumerate}

\section{Remarks}
SUSY-QM has allowed us to show that new classes of exact solutions are found
for the WDW equation, extending the class of exactly solvable spectral 
problem for the 
Schr\"odinger like hamiltonian in one space dimension (in quantum cosmology
the WDW equation plays this role).
In this work, we found isospectral cosmological potential and a one-parameter
family of wavefunctions of the universe for  barotropic FRW models with a 
cosmological constant, including the
factor ordering problem. We were able to find exact particular solutions for
different choices of the $\gamma$ parameter in the WDW equation (\ref{WDW})
and different universes. 
The parameter $\rm \lambda$ looks like a 
decoherence parameter embodying a sort of quantum cosmological dissipation
(or damping) distance (see Fig. (\ref{u-1})).

\end{document}